\documentclass[10pt,pra,aps,notitlepage,superscriptaddress,reprint]{revtex4-2}

\usepackage[hyperindex,breaklinks,colorlinks=true]{hyperref}

\usepackage{bm}

\usepackage[utf8]{inputenc}
\usepackage[T1]{fontenc}
\usepackage[english]{babel}  

\usepackage{tikz}

\usepackage{colortbl}
\usepackage{tabu,diagbox}

\usepackage{pgfplots}
\usepackage{lipsum}
\usepackage{youngtab}
\usepackage{subfigure}

\usepackage{times}
\usepackage{dsfont}

\usepackage{amssymb,amsmath,amsfonts,amsthm}
\usepackage{mathtools}
\usepackage{verbatim}
\usepackage{enumerate}
\usepackage{graphicx}

\usepackage[capitalise]{cleveref}
\usepackage{bbm}
\usepackage{booktabs}
\usepackage{hyperref}

\newcommand{\appsection}[1]{\let\oldthesection\thesection
  \renewcommand{\thesection}{Appendix \oldthesection}
  \section{#1}\let\thesection\oldthesection}

\def\Dbar{\leavevmode\lower.6ex\hbox to 0pt
{\hskip-.23ex\accent"16\hss}D}

\newcommand{\nc}{\newcommand}

\nc{\cA}{{\cal A}} \nc{\cB}{{\cal B}} \nc{\cC}{{\cal C}}
\nc{\cD}{{\cal D}} \nc{\cE}{{\cal E}} \nc{\cF}{{\cal F}}
\nc{\cG}{{\cal G}} \nc{\cH}{{\cal H}} \nc{\cI}{{\cal I}}
\nc{\cJ}{{\cal J}} \nc{\cK}{{\cal K}} \nc{\cL}{{\cal L}}
\nc{\cM}{{\cal M}} \nc{\cN}{{\cal N}} \nc{\cO}{{\cal O}}
\nc{\cP}{{\cal P}} \nc{\cQ}{{\cal Q}} \nc{\cR}{{\cal R}}
\nc{\cS}{{\cal S}} \nc{\cT}{{\cal T}} \nc{\cU}{{\cal U}}
\nc{\cV}{{\cal V}} \nc{\cW}{{\cal W}} \nc{\cX}{{\cal X}}
\nc{\cZ}{{\cal Z}}

\def\affa{\affiliation{Shenzhen SpinQ Technology Co., Ltd., Shenzhen, China}}
\def\affb{\affiliation{College of Physics and Electronic Engineering \& Center for Computational Sciences,  Sichuan Normal University, Chengdu, China}}
\def\affc{\affiliation{Department of Physics, The Hong Kong University of Science and Technology, Clear Water Bay, Kowloon, Hong Kong}}

\def\afff{\affiliation{Institute for Quantum Computing, University of Waterloo, Waterloo, Ontario, Canada}}

\begin{document}

\title{SpinQ Triangulum: a commercial three-qubit desktop quantum computer}

\author{Guanru Feng}
\email{gfeng@spinq.cn}
\affa

\author{Shi-Yao Hou}
\affb
\affa

\author{Hongyang Zou}
\affa

\author{Wei Shi}
\affa

\author{Sheng Yu}
\affa 

\author{Zikai Sheng}
\affa 

\author{Xin Rao}
\affa 

\author{Kaihong Ma}
\affa 

\author{Chenxing Chen}
\affa

\author{Bing Ren}
\affa

\author{Guoxing Miao}
\afff
\affa

\author{Jingen Xiang}
\email{jxiang@spinq.cn}
\affa

\author{Bei Zeng} 
\email{zengb@ust.hk}
\affc

\begin{abstract}

SpinQ Triangulum is the second generation of the desktop quantum computers designed and manufactured by SpinQ Technology. SpinQ's desktop quantum computer series, based on room temperature NMR spectrometer, provide light-weighted, cost-effective 
and maintenance-free quantum computing platforms that aim to provide real-device experience for quantum computing education for K-12 and college level. These platforms also feature quantum control design capabilities for studying quantum control and quantum noise.
Compared with the first generation product, the two-qubit SpinQ Gemini, Triangulum features a three-qubit QPU, smaller dimensions ($61 \times 33 \times 56$ cm$^3$) and lighter (40 kg). Furthermore, the magnetic field is more stable and the performance of quantum control is more accurate. This paper introduces the system design of Triangulum and its new features. As an example of performing quantum computing tasks, we present the implementation of the Harrow-Hassidim-Lloyd (HHL) algorithm on Triangulum, demonstrating Triangulum's capability of undertaking complex quantum computing tasks. SpinQ will continue to develop desktop quantum computing platform with more qubits. Meanwhile, a simplified version of SpinQ Gemini, namely Gemini Mini \footnote{\url{https://www.spinq.cn/products\#geminiMini-anchor}}
, has been recently realised. Gemini Mini is much more portable ($20\times 35
\times 26$ cm$^3$, 14 kg) and affordable for most K-12 schools around the world.

\end{abstract}

\date{\today}


\maketitle

\section{Introduction}

\begin{figure*}[!htbp]
\includegraphics[scale=0.77]{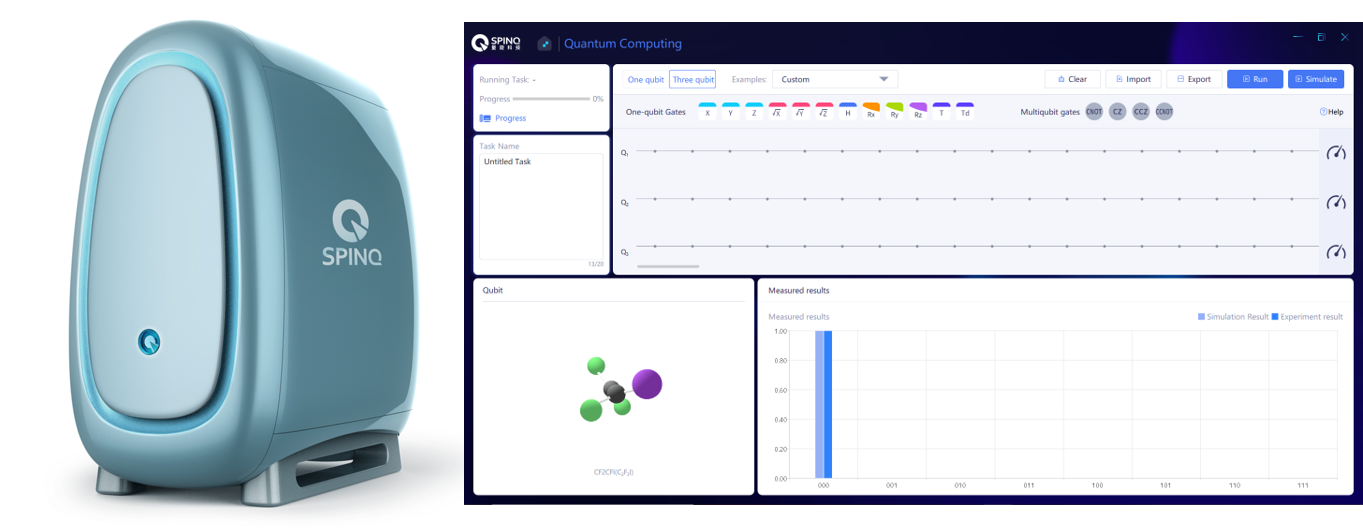}
\caption{(a) The photo of Triangulum and (b) User-friendly interface  \textsc{SpinQuasar}~\cite{spinqweb}. \textsc{SpinQuasar} is installed on a PC that connects with Triangulum. Users can manipulate Triangulum conveniently via \textsc{SpinQuasar}. On this particular page, users can compose quantum circuits, implement quantum algorithms, and check the computation results. There are two buttons, 'Run' and 'Simulate', for activation of the experiment and the simulation, respectively. The probabilities of the eight eigen base from the experiment and the simulation are shown in the bottom half of the interface.}
\label{photo}
\end{figure*}

SpinQ Triangulum is the second generation product of SpinQ's commercial desktop quantum computing platform series~\cite{spinqweb,patent1,Gemini}. Similar to the first generation product, the two-qubit SpinQ Gemini, Triangulum is based on nuclear magnetic resonance (NMR) system, which was among the very 
first systems developed for quantum computing and has developed many advanced quantum control techniques~\cite{Corypps,Gershenfeld97bulkspin-resonance,ChuangBulk,CORY199882,Rayonequbit,Cory2000,LINDEN199861,nmr1,nmr2,nmr3,nmr4,nmr5,nmr6,Fourier,
ShorExp,ErrorCorrection,ErrorCorrection5,
geometricNMR,adiabaticNMR,CP,CPMG,strongly,compositeChuang,grape1,nmrrev2005,compositeJones,RobustDDSuter}.
Compared with Gemini,
Triangulum accommodates a three-qubit QPU. The weight and size are further reduced to $40$ kg and $61 \times 33 \times 56$ cm$^3$. Fig. \ref{photo} shows the exterior look of Triangulum and its user-friendly interface \textsc{SpinQuasar}, where users can compose quantum circuits and interact with the QPU. 

Most of the quantum computers in research labs are out of reach in real life, due to their cost, weight, volume and extreme physical conditions, such as traditional NMR quantum computing which is performed on huge and expensive commercial superconducting NMR spectrometers, and superconducting qubit quantum computing that requires extreme temperature environment. SpinQ desktop quantum computing platforms take advantage of the recent development of small permanent magnet technology~\cite{magnet1} to reach a small size and weight. SpinQ Gemini is only 55 kg and Triangulum is only 40 kg. Furthermore, compared with their counterparts in research labs, SpinQ Gemini and Triangulum are cost-effective, and require no special maintenance, hence very friendly to everyone who is interested in quantum computing. 

SpinQ Triangulum inherits Gemini's powerful functions of  quantum algorithm circuit design and programming using its software \textsc{SpinQuasar} (Fig.~\ref{photo}), as well as the demonstrations of $10+$ famous quantum algorithms, such as Deutsch algorithm~\cite{DJalgorithm} and Grover algorithm~\cite{grover1996fast,Groverlong}. Thus it provides a very friendly platform for non-specialists who aim to learn quantum computing basics and quantum programming. Furthermore, Triangulum has improved its stability and quantum control accuracy. Its powerful arbitrary-waveform generation function enables advanced control of the quantum system. With its three-qubit QPU, Triangulum can serve as a powerful tool for quantum computing related research under real world conditions. 

In this paper, we introduce the system of SpinQ Triangulum. In Sec. II, we briefly introduce the hardware and software design. In Sec. III, we introduce how quantum computing is fulfilled on Triangulum. In Sec. IV, We demonstrate the implementation of Harrow-Hassidim-Lloyd (HHL) quantum algorithm on Triangulum. A discussion on future plans of next generations products will follow in Sec. V.

\section{System}
\begin{figure*}[!htb]
\includegraphics[scale=0.45]{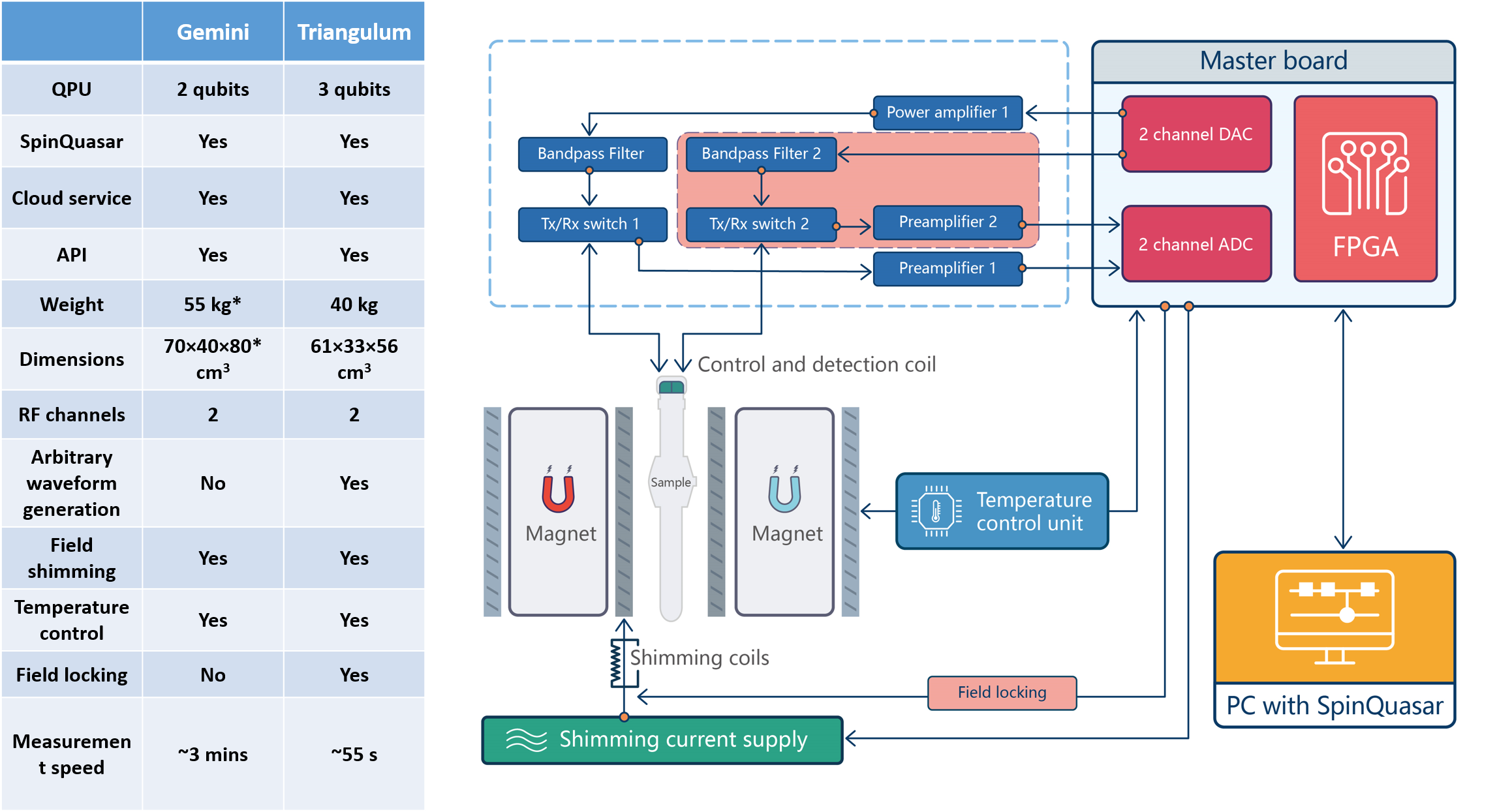}
\caption{Left: The comparison between Gemini and Triangulum. *These are the data of the first version of Gemini. The weight and dimensions of the current version of Gemini are similar to those of Triangulum.   Right: The overview of the schematic diagram of Triangulum system. The master board equipped with an FPGA, provides the control logic of Triangulum. \textsc{SpinQuasar} communicates with FPGA through USB so that the user can access Triangulum. The magnets, together with the temperature control unit, the field shimming system and the field locking system provide a stable static homogeneous magnetic field. The RF module amplifies the RF control pulses and detects the RF signals from the qubits. The field locking system utilizes the RF signal route that is composed of {\bf Bandpass Filter 2}, {\bf Tx/Rx switch 2}, {\bf Preamplifier 2}, which are in the orange boxs.}
\label{hardware}
\end{figure*}

The overall schematic diagram is shown in Fig.~\ref{hardware}.
Similar to Gemini, Triangulum is composed of  a PC with \textsc{SpinQuasar}, a control system on the master board, a temperature control module, a pair of permanent magnets, a field shimming system, a radio frequency (RF) system, and  a tube of sample. Additionally, Triangulum has a field locking system, and its pulse generator enables arbitrary waveform generation function. 

The modules that Triangulum inherits from Gemini realize most of the same functions\cite{Gemini}. The master board includes an FPGA, an analog-digital converter (ADC) and a digital-analog converter (DAC), which altogether realize the algorithms required for pulse generation, signal processing, etc. Triangulum's pulse generation function is more powerful than Gemini with arbitrary waveform generation ability, which will be discussed later. \textsc{SpinQuasar} is the software interface for users, which provides an interface to the QPU as well as the instrument calibration. Advanced functions such as cloud computing~\cite{spinqCloud}, and APIs for programmable control, are also supported by Triangulum.

The NMR sample used in Triangulum is iodotrifluoroethylene (C$_2$F$_3$I). Different from Gemini which uses two different types of nuclei as qubits, Triangulum uses three $^{19}$F nuclei as three qubits. On one hand, only one RF channel is needed for excitation and signal detection for the three qubits, which seems to be a great advantage. On the other hand, simple square pulses used on Gemini are not enough anymore to manipulate nuclei of the same type as different qubits. Pulses with arbitrary shapes, such as the GRadient Ascent Pulse Engineering (GRAPE) pulses \cite{grape1}, are needed for accurate quantum control. Therefore the control of three qubits of the same type spins requires the pulse generator and the RF amplification/transmission system to be more powerful. In Triangulum the pulse generation module of FPGA is developed to be capable of arbitrary waveform generation, with the magnitude accuracy of 1/65536 and phase accuracy of 2$\pi$/65536.  The RF system which is responsible for pulse amplification and transmission, is also improved to faithfully transmit arbitrary waveforms.

To realize a stable and homogeneous magnetic field required by quantum computing, Gemini 
uses a pair of NdFeB plate permanent magnets, a field shimming system and a temperature control system. In addition to those modules, Triangulum is equipped with a field locking system to make the static magnetic field more stable. The increased field stability satisfies the requirement  of specially designed pulse shapes that are very sensitive to the resonance frequency fluctuations of qubits and thus very sensitive to the field fluctuation. Locking is realized by continuously exciting and detecting $^{1}$H signals of acetone. By analyzing the detected $^{1}$H frequency, the field drift can be estimated. Then a compensation field is generated by coils to make the magnet field stable at a desired magnitude. The $^{1}$H spin excitation and signal detection are realized using the second RF channel as shown  in Fig. \ref{hardware}.

Therefore, with the additional abilities, field locking and arbitrary  waveform generation, the three-qubit system of Triangulum can be accurately manipulated for different quantum computing tasks.

\begin{figure}[!ht]
\includegraphics[width=8cm]{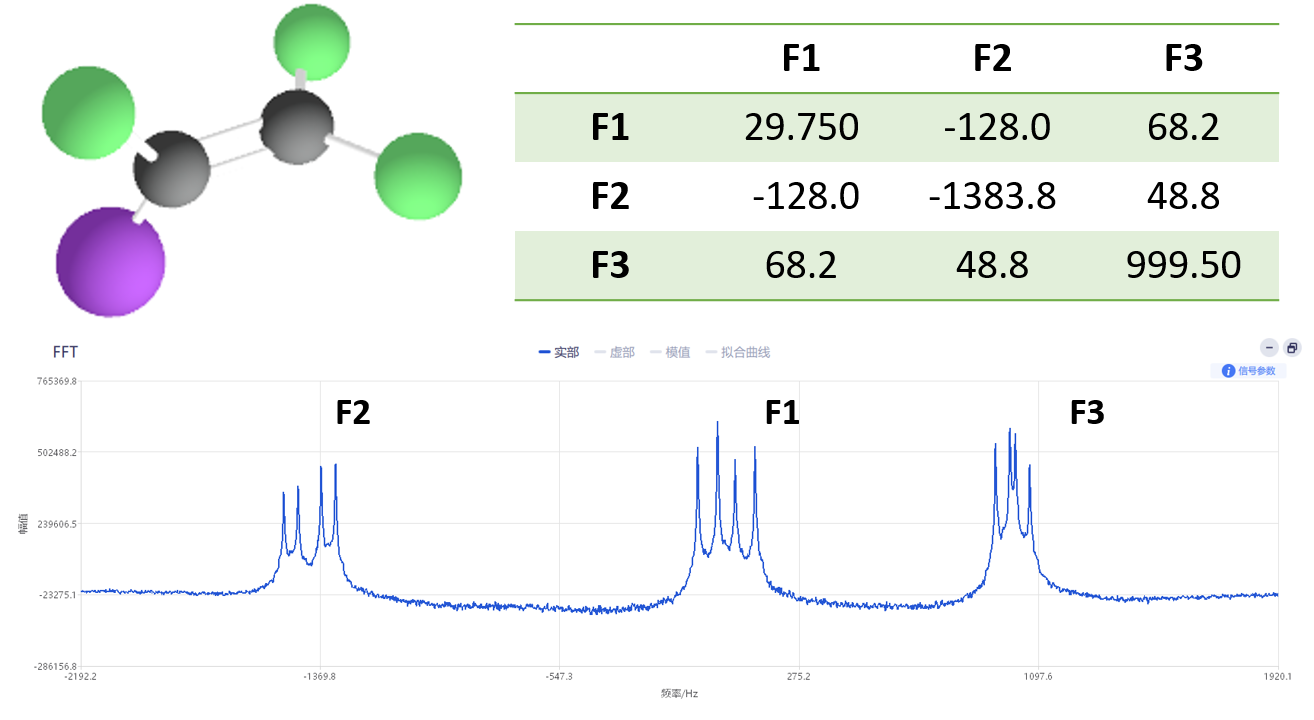}
\caption{The molecule structure (left), its parameter table (right), the Fourier transform spectrum of $^{19}$F (bottom). In the molecule, there are three $^{19}$F nuclear spins (green). The diagonal elements of the table are the chemical shifts of the three $^{19}$F. The off diagonal elements are their J couplings. All the values in the table are in unit of Hz. Each of the $^{19}$F spins has four peaks in the Fourier transform spectrum, with peak splittings determined by the J coupling constants.}
\label{molecule}
\end{figure}

\section{Quantum computation}

\subsection{Software interface}

Same as Gemini, \textsc{SpinQuasar} is provided to users as the interface to Triangulum (\cref{photo}). Most parts of the interface are same as  the version on Gemini. The quantum circuits as well as the available quantum gates are shown in the top half of the interface. A drop-down list provides users with built-in algorithms. The 'Run' and 'Simulate' buttons on the right of the circuits can be used to initiate the experiments on the quantum processor or the embedded quantum simulator. The results are given at the bottom right. It should be mentioned that, users can click on or off the measurement labels to the rightmost of the lines which stand for qubits. The experiment or simulation will only give the measurement results for the qubits which users choose. In the following part, we will introduce the molecule system of the quantum processor, the available quantum gates, the initial state and the measurement of the quantum processor state.

\subsection{The spin system}
The sample we use is iodotrifluoroethylene (C$_2$F$_3$I). The molecules are placed in the center of the parallel permanent magnets. A $^{19}$F nucleus is a spin-half system. When placed in a static magnetic field, it has two energy levels, thus a $^{19}$F spin can be used as a qubit. The Lamor frequency of  $^{19}$F in 1 T magnetic field is $\sim$40 MHz. Its state can be manipulated by irradiating electro-magnetic waves (pulses) with frequencies close to its Larmor frequency. The three $^{19}$F nuclei in iodotrifluoroethylene are used as the three qubits. The structure and the parameters of the sample are listed in Fig. \ref{molecule}.
The $T_1$ and $T_2$ for the $^{19}$F spins are about 7s and 0.2s, respectively. The J couplings between the three spins are  -128 Hz, 68 Hz and 49 Hz. The frequencies of the three $^{19}$F spins, located around 40 MHz, are slightly different. These differences are usually called chemical shifts. The excitation profile in the frequency domain of simple square pulses is usually broad. Because of the small frequency difference between the three spins, it is difficult to realize accurate individual controls using simple square pulses. Therefore, GRAPE pulses are used to control the three-qubit system. The spin Hamiltonian in the rotating frame is 
\begin{align} 
H_0&=2\pi (\nu_1 I_z^{1}+\nu_2 I_z^{2}+\nu_3 I_z^{3})\nonumber\\
&+ 2\pi (J_{1,2} I_z^{1}I_z^{2}+J_{1,3} I_z^{1}I_z^{3}+J_{2,3} I_z^{2}I_z^{3}),
\end{align}
where $\nu_i$'s are chemical shifts for the $i$th nuclei and  $J_{i,j}$'s  are J couplings between the $i$th and $j$th nuclei.

\subsection{The gate set}
The global 90 degree rotation of the three qubits can be realized using 10 us square pulses. However, global control is not enough in quantum computing. As mentioned above, on Triangulum GRAPE pulses are used to realize all the quantum gates available. The available quantum gates contain single-qubit, two-qubit and three-qubits gates. The single-qubit gates are Pauli gates ($\sigma_{x}$, $\sigma_{y}$, $\sigma_{z}$), 90 degree rotation gates and arbitrary rotation gates along $x$, $y$ and $z$ axes, Hadamard gates, T gates and inverse of T gates. The two-qubit gates are CNOT and CZ gates between any pairs of the three qubits. The three-qubit gates are the Toffli gates with any two of the three qubits as the control qubits, and the CCZ gates which implement a $\pi$ phase change to $|111\rangle$. It should be mentioned that all the gates are realized using a single GRAPE pulse, except the following cases (a) all $z$ rotation gates, including $\sigma_{z}$, 90 degree rotation and arbitrary angle rotation gates along $z$ axis, (b) arbitrary rotation gates along $x$ and  $y$ axes.

The case (a), $z$ rotation gates, are realized virtually by changing the phase of the reference rotating frame\cite{PhysRevA.78.012328}. To illustrate how this works, we first consider a simple square rotation pulse, with rotating axis $\phi$,
\begin{align}
U=e^{-it\Omega(\cos\phi\sigma_x/2+\sin\phi\sigma_y/2)}.
\end{align}
$\Omega$ and $\phi$ are the pulse amplitude and phase, respectively, which can be adjusted by the arbitrary waveform generator. For any experiment, a reference phase $\phi_0$ is set for observation so that all the pulse phase is relative to $\phi_0$. 
\begin{align}
U=e^{-it\Omega(\cos(\phi-\phi_0)\sigma_x/2+\sin(\phi-\phi_0)\sigma_y/2)}.
\end{align}
Changing $\phi-\phi_0$ means rotating the rotation axis in the $x$-$y$ plane around $z$ axis. If within a experiment, we change the reference phase $\phi_0$,  it means the reference frame is rotated along $z$ axis. For a given quantum state, this reference frame change is equivalent to a $z$ rotation. Thus by changing the reference frame of pulses, which can be done conveniently in observation, one can realize virtual $z$ rotations. And within a experiment, after any virtual $z$ rotations, since the reference frame is changed, $\phi$ of the subsequent rotations should be adjusted accordingly for desired rotation axis. Here we use the following sequence as an example.
\begin{align}
R_x(\theta)-R_z(\frac{\pi}{2})-R_x(\gamma).
\end{align}
We rotate the reference frame by $-\pi/2$ to realize $R_z(\frac{\pi}{2})$ virtually. This means we need to change the observing frame by $-\pi/2$. And the sequence that needs to be implemented  is
\begin{align}
R_x(\theta)-R_{-y}(\gamma).
\end{align}
The $R_x(\gamma)$ after the virtual $z$ rotation is changed accordingly to  $R_{-y}(\gamma)$.

The gates in case (b), the arbitrary $x$ or $y$ rotation gates, are realized by first rotating the $x$ or $y$ axes to $z$ using GRAPE pulses, then implementing a $z$ rotation and finally rotating the axes back to $x$ or $y$.  

\subsection{The pseudo-pure state}
The initial state of the three-qubit system is prepared to be a pseudo-pure state (PPS) \cite{Corypps} using the same method as in Gemini. 

A PPS  has the following form,
\begin{align}
\rho_{\mathrm{pps}}=\frac{1-\eta}{2^n}I^{\otimes n}+\eta|\psi\rangle\langle\psi|.
\end{align}
$|\psi\rangle$ is a pure state. A PPS above has the same unitary dynamics and observable effects as the pure state $|\psi\rangle$ except for the factor $\eta$. PPS is widely used in NMR quantum computation. 

Triangulum utilizes the relaxation method in Ref. \cite{pps} to prepare the three-qubit PPS starting from the thermal equilibrium state, whose state population is subject to Boltzmann distribution at room temperature. The relaxation method uses repetitions of basis permutation operations and T1 relaxation. Different from Gemini, whose permutation operation is realized by square pulses and delays, on Triangulum the basis permutation gate,
\begin{align}
&U_{\mathrm{permute}}=\begin{pmatrix}
1 & 0 & 0 & 0 & 0 & 0 & 0& 0\\
0 & 0 & 0 & 0 & 0 & 0 & 0& 1\\
0 & 1 & 0 & 0 & 0 & 0 & 0& 0\\
0 & 0 & 1 & 0 & 0 & 0 & 0& 0\\
0 & 0 & 0 & 1 & 0 & 0 & 0& 0\\
0 & 0 & 0 & 0 & 1 & 0 & 0& 0\\
0 & 0 & 0 & 0 & 0 & 1 & 0& 0\\
0 & 0 & 0 & 0 & 0 & 0 & 1& 0\\
\end{pmatrix},
\end{align}
is realized by a GRAPE pulse. $U_{\mathrm{permute}}$ permutes the state population between basis $|001\rangle, |010\rangle, |011\rangle, |100\rangle, |101\rangle, |110\rangle, |111\rangle$ while leaving $|000\rangle$ unchanged. This permutation operation is combined with a delay after it. The permutation and the T1 relaxation in the delay take effect alternately. After a certain number of cycles, the system can reach a state whose dominantly occupied basis is $|000\rangle$ and the other seven base have the same but smaller probability. This obtained state is a PPS and can be used as the initial state $|000\rangle$. 
\subsection{Measurements}
Different from Gemini, Triangulum only measures the diagonal elements of the system's density matrix after a certain gate sequence is applied. This means only probabilities of $|000\rangle, |001\rangle, |010\rangle, |011\rangle, |100\rangle, |101\rangle, |110\rangle, |111\rangle$ are measured. In NMR quantum computing, results are collected by ensemble measurements, which means one collects the ensemble expectation values of a certain operator. Specifically, NMR collects the $\langle \sigma_x+i\sigma_y\rangle$ of a certain qubit. To obtain the diagonal elements of the system's density matrix, one needs to measure $\langle \sigma_z^1\rangle$, $\langle \sigma_z^2\rangle$, $\langle \sigma_z^3 \rangle$, $\langle \sigma_z^1\sigma_z^2 \rangle$, $\langle \sigma_z^1\sigma_z^3 \rangle$, $\langle \sigma_z^2\sigma_z^3 \rangle$ , $\langle \sigma_z^1\sigma_z^2\sigma_z^3 \rangle$.  The following expression illustrates how the 8 diagonal elements of the density matrix are calculated from those expectation values
\begin{align}
\rho_{11}=&\frac{1}{8}(1+\langle \sigma_z^1\rangle+\langle \sigma_z^2\rangle+\langle \sigma_z^3 \rangle+\langle \sigma_z^1\sigma_z^2 \rangle\nonumber\\&+\langle \sigma_z^1\sigma_z^3 \rangle+\langle \sigma_z^2\sigma_z^3 \rangle+\langle \sigma_z^1\sigma_z^2\sigma_z^3 \rangle),\\
\rho_{22}=&\frac{1}{8}(1+\langle \sigma_z^1\rangle+\langle \sigma_z^2\rangle-\langle \sigma_z^3 \rangle+\langle \sigma_z^1\sigma_z^2 \rangle\nonumber\\&-\langle \sigma_z^1\sigma_z^3 \rangle-\langle \sigma_z^2\sigma_z^3 \rangle-\langle \sigma_z^1\sigma_z^2\sigma_z^3 \rangle),\\
\rho_{33}=&\frac{1}{8}(1+\langle \sigma_z^1\rangle-\langle \sigma_z^2\rangle+\langle \sigma_z^3 \rangle-\langle \sigma_z^1\sigma_z^2 \rangle\nonumber\\&+\langle \sigma_z^1\sigma_z^3 \rangle-\langle \sigma_z^2\sigma_z^3 \rangle-\langle \sigma_z^1\sigma_z^2\sigma_z^3 \rangle),\\
\rho_{44}=&\frac{1}{8}(1+\langle \sigma_z^1\rangle-\langle \sigma_z^2\rangle-\langle \sigma_z^3 \rangle-\langle \sigma_z^1\sigma_z^2 \rangle\nonumber\\&-\langle \sigma_z^1\sigma_z^3 \rangle+\langle \sigma_z^2\sigma_z^3 \rangle+\langle \sigma_z^1\sigma_z^2\sigma_z^3 \rangle),\\
\rho_{55}=&\frac{1}{8}(1-\langle \sigma_z^1\rangle+\langle \sigma_z^2\rangle+\langle \sigma_z^3 \rangle-\langle \sigma_z^1\sigma_z^2 \rangle\nonumber\\&-\langle \sigma_z^1\sigma_z^3 \rangle+\langle \sigma_z^2\sigma_z^3 \rangle-\langle \sigma_z^1\sigma_z^2\sigma_z^3 \rangle),\\
\rho_{66}=&\frac{1}{8}(1-\langle \sigma_z^1\rangle+\langle \sigma_z^2\rangle-\langle \sigma_z^3 \rangle-\langle \sigma_z^1\sigma_z^2 \rangle\nonumber\\&+\langle \sigma_z^1\sigma_z^3 \rangle-\langle \sigma_z^2\sigma_z^3 \rangle+\langle \sigma_z^1\sigma_z^2\sigma_z^3 \rangle),\\
\rho_{77}=&\frac{1}{8}(1-\langle \sigma_z^1\rangle-\langle \sigma_z^2\rangle+\langle \sigma_z^3 \rangle+\langle \sigma_z^1\sigma_z^2 \rangle\nonumber\\&-\langle \sigma_z^1\sigma_z^3 \rangle-\langle \sigma_z^2\sigma_z^3 \rangle+\langle \sigma_z^1\sigma_z^2\sigma_z^3 \rangle),\\
\rho_{88}=&\frac{1}{8}(1-\langle \sigma_z^1\rangle-\langle \sigma_z^2\rangle-\langle \sigma_z^3 \rangle+\langle \sigma_z^1\sigma_z^2 \rangle\nonumber\\&+\langle \sigma_z^1\sigma_z^3 \rangle+\langle \sigma_z^2\sigma_z^3 \rangle-\langle \sigma_z^1\sigma_z^2\sigma_z^3 \rangle),
\label{qstdecomp}
\end{align}
It should be noted that all the $\langle \sigma_z^1\rangle$, $\langle \sigma_z^2\rangle$, $\langle \sigma_z^3 \rangle$, $\langle \sigma_z^1\sigma_z^2 \rangle$, $\langle \sigma_z^1\sigma_z^3 \rangle$, $\langle \sigma_z^2\sigma_z^3 \rangle$ , $\langle \sigma_z^1\sigma_z^2\sigma_z^3 \rangle$ values are not direct observable. Additional readout pulses are needed to transform them to be observable. Totally three experiments are needed to obtain the above six values. Specifically, in the three experiments, each of the three qubits are measured after a 90 degree readout pulse on that particular qubit, respectively. In each experiment, four peaks of the measured qubit are obtained in the spectrum, whose amplitudes are linear combinations of $\langle \sigma_z^1\rangle$, $\langle \sigma_z^2\rangle$, $\langle \sigma_z^3 \rangle$, $\langle \sigma_z^1\sigma_z^2 \rangle$, $\langle \sigma_z^1\sigma_z^3 \rangle$, $\langle \sigma_z^2\sigma_z^3 \rangle$ , $\langle \sigma_z^1\sigma_z^2\sigma_z^3 \rangle$. For example, in one of the three experiments, the first qubit is observed after a $R_y^{1}(\pi/2)$ rotation. Then the real parts of the four peaks of the first qubit in the Fourier transform spectrum are proportional to the following values
\begin{align}
\langle \sigma_z^1\rangle-\langle \sigma_z^1\sigma_z^3\rangle+\langle \sigma_z^1\sigma_z^2\rangle-\langle \sigma_z^1\sigma_z^2\sigma_z^3\rangle,\nonumber\\
\langle \sigma_z^1\rangle+\langle \sigma_z^1\sigma_z^3\rangle+\langle \sigma_z^1\sigma_z^2\rangle+\langle \sigma_z^1\sigma_z^2\sigma_z^3\rangle,\nonumber\\
\langle \sigma_z^1\rangle-\langle \sigma_z^1\sigma_z^3\rangle-\langle \sigma_z^1\sigma_z^2\rangle+\langle \sigma_z^1\sigma_z^2\sigma_z^3\rangle,\nonumber\\
\langle \sigma_z^1\rangle+\langle \sigma_z^1\sigma_z^3\rangle-\langle \sigma_z^1\sigma_z^2\rangle-\langle \sigma_z^1\sigma_z^2\sigma_z^3\rangle.
\end{align}
By fitting the spectra and solving linear equations,  $\langle \sigma_z^1\rangle$, $\langle \sigma_z^2\rangle$, $\langle \sigma_z^3 \rangle$, $\langle \sigma_z^1\sigma_z^2 \rangle$, $\langle \sigma_z^1\sigma_z^3 \rangle$, $\langle \sigma_z^2\sigma_z^3 \rangle$ , $\langle \sigma_z^1\sigma_z^2\sigma_z^3 \rangle$ can be obtained and thus the $\rho_{ii}$'s can be calculated for $i$ from 1 to 8. It only takes $\sim$55 s for Triangulum to reconstruct the diagonal elements of the density matrix, faster than Gemini which needs six experiments for density matrix reconstruction.

\begin{figure}[!ht]
\includegraphics[width=8cm]{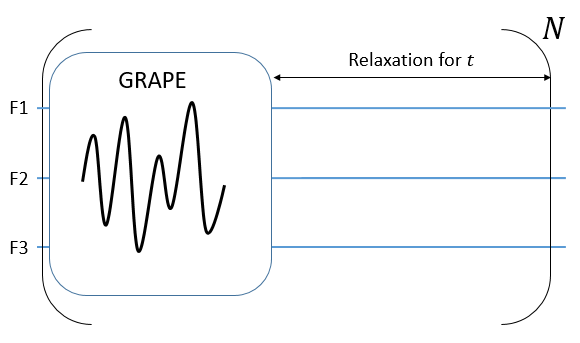}
\caption{The pulse sequence for pseudo-pure state preparation. The first permutation gate is realized using a GRAPE pulse. After it is a long delay within which the natural relaxation takes effect. By properly choosing the repetition number, $N$, and the duration of the delay, $t$, the system can be steered to the pseudo-pure state $|000\rangle$ from the thermal equilibrium state.  }
\label{pps}
\end{figure}

\section{Application: HHL algorithm for linear systems of equations}
In this section, we demonstrate the realization of the Harrow-Hassidim-Lloyd (HHL) algorithm for linear systems of equations on Triangulum.

Solving linear systems of equations is a problem present in almost all areas of science and engineering. HHL algorithm \cite{PhysRevLett.103.150502} is a quantum algorithm for solving linear systems of equations. Under certain conditions, this algorithm has an exponential speedup over the fastest classical algorithm. Thus there are attractive potential applications of HHL algorithm as the data size is ever growing and solving linear systems of equations is more and more demanding in science and engineering. HHL algorithm has become a subroutine in many quantum algorithms. For example, many quantum machine learning algorithms make use of HHL algorithm. Here we implement a simplified version of HHL algorithm \cite{HHL3}.

\subsection{HHL algorithm}
The linear system of equations we are trying to solve is
\begin{align}
A\vec{x}=\vec{b},
\end{align}
where $A$ is an $N\times N$ Hermitian matrix,  $\vec{b}$ is a known $N\times 1$ vector, and $\vec{x}$, an unknown $N\times 1$ vector, is to be solved. To use HHL algorithm, the first step is to express $\vec{x}$ and $\vec{b}$ in the form of quantum states, $|x\rangle$ and $|b\rangle$, which are normalized and have vectors proportional to $\vec{x}$ and $\vec{b}$, respectively. $A$ can be considered as the matrix form of an operator, $\hat{A}$. Therefore, solving $\vec{x}=A^{-1}\vec{b}$ has been mapped to solving $|x\rangle\propto\hat{A}^{-1}|b\rangle$. $\hat{A}$ can be expressed as $\Sigma_i \lambda_i|u_i\rangle\langle u_i|$, where $\lambda_i$ are eigen values of $\hat{A}$ and $|u_i\rangle$ are the corresponding eigen states. $|b\rangle$ can be expressed using $|u_i\rangle$, $|b\rangle=\Sigma_i\beta_i|u_i\rangle$. Therefore, $|x\rangle$ can be written as 
\begin{align}
|x\rangle\propto\hat{A}^{-1}|b\rangle=\Sigma_i\frac{\beta_i}{\lambda_{i}}|u_i\rangle.
\end{align}
HHL algorithm utilizes a composite system with three subsystems to derive $\Sigma_i\frac{\beta_i}{\lambda_{i}}|u_i\rangle$. The first subsystem contains ${\rm log}_2N$ qubits to store $|b\rangle$ and the final result $|x\rangle$. The second subsystem is an $n$-qubit register used to derive the eigen values of $\hat{A}$. The third subsystem is an ancilla qubit that assists in deriving the reciprocal of $\lambda_i$.
\begin{figure}[!ht]
\includegraphics[width=9cm]{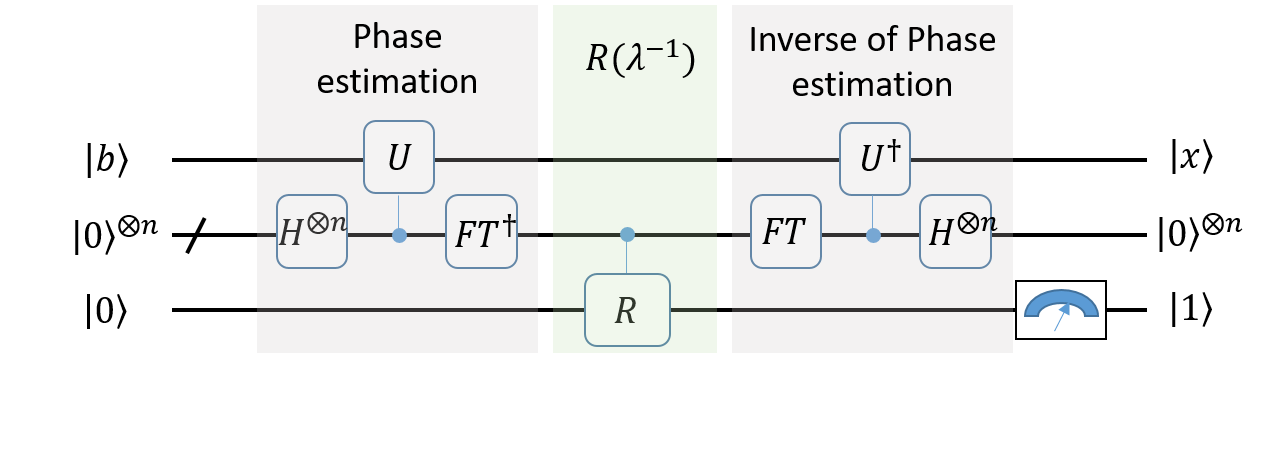}
\caption{The general circuit for HHL algorithm.}
\label{HHL}
\end{figure}

There are three steps in HHL algorithm, as shown in Fig. \ref{HHL} . The system is initialized in the state
\begin{align}
|b\rangle|0\rangle^{\otimes n}|0\rangle=\Sigma_i\beta_i|u_i\rangle|0\rangle^{\otimes n}|0\rangle.
\end{align}
The first step is to implement the quantum phase estimation algorithm to estimate the eigen values of $\hat{A}$ with an accuracy of $n$-bits. In this step, the required controlled-$U$ operation realizes $e^{ik\hat{A}t_0/2^n}$ on the first subsystem when the $n$-qubit register is in the state $|k\rangle$, where $|k\rangle$ is the quantum state corresponding to the $n$-bit binary form of $k$, and $t_0$ is usually chosen to be $2\pi$. After this, the state of the whole system is $\Sigma_i\beta_i|u_i\rangle|\lambda_i\rangle|0\rangle$, where $|\lambda_i\rangle$ is the quantum state corresponding to the $n$-bit binary form of $\lambda_i$.

The second step is a controlled rotation of the third subsystem. The $n$-qubits in the register are used as the control qubits. When they are in the state $|\lambda_i\rangle$, the third subsystem is rotated by an angle $2\sin^{-1}(C/\lambda_i)$ (here $C$ is a properly chosen constant). After this, the state of the whole system is
\begin{align}
\Sigma_i\beta_i|u_i\rangle|\lambda_i\rangle(\sqrt{1-|\frac{C}{\lambda_i}|^2}|0\rangle+\frac{C}{\lambda_i}|1\rangle).
\end{align} 

The last step is the reverse of phase estimation. After this step, the $n$-qubit register is disentangled. The state of the whole system is
\begin{align}
\Sigma_i\beta_i|u_i\rangle|0\rangle^{\otimes n}(\sqrt{1-|\frac{C}{\lambda_i}|^2}|0\rangle+\frac{C}{\lambda_i}|1\rangle).
\end{align}
Now if the third subsystem is measured and the result is $|1\rangle$, then the first subsystem is in the state of $\Sigma_i\frac{C}{\lambda_i}\beta_i|u_i\rangle$ which is proportional to $|x\rangle$ and thus is the solution. The probability to get $|1\rangle$ of the third subsystem is $\Sigma_i|\frac{C\beta_i}{\lambda_i}|^2$, and this is also the probability of success.

It should be mentioned that the original paper of HHL algorithm does not give a detailed method to implement the controlled rotation $R(\lambda^{-1})$ in the second step. The controlled unitary in the step of quantum phase estimation can be intuitively decomposed to individual one-qubit-controlled operations using each of the register qubits as the control qubit. It is not the case for the controlled rotation $R(\lambda^{-1})$. The authors in Ref. \cite{doi:10.1080/00268976.2012.668289} proposed a way to realize the controlled rotation $R(\lambda^{-1})$ by introducing additional ancilla qubits. Here we will not go to the details of their method. In the three-qubit case, the $R(\lambda^{-1})$ operation can be realized in a simple way which will be described in the following section.

\subsection{HHL algorithm simplified in three-qubit case} 
In a three-qubit system, we will use each qubit as one of the three subsystems required by HHL algorithm. The fact that the first subsystem has one qubit means $A$ is a $2\times 2$ matrix with two eigen values. That the second subsystem has one qubit means, when estimating the eigen values of $A$ the accuracy is only one bit. In the step of quantum phase estimation, the required quantum Fourier transformation (FT) can be realized  using only one Hadamard gate, $H$.

\begin{figure}[!ht]
\includegraphics[width=9cm]{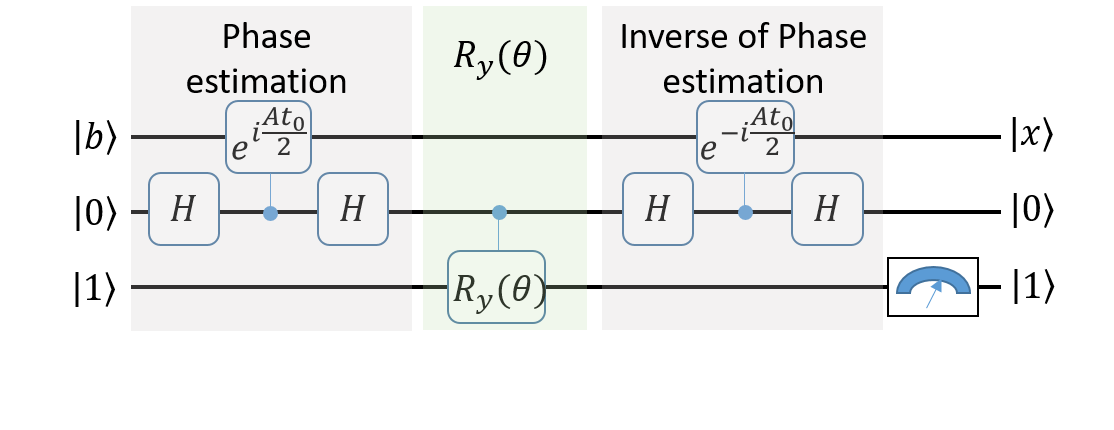}
\caption{The HHL circuit in the three-qubit case.}
\label{HHL1}
\end{figure}

With only one register qubit, the controlled operation in phase estimation can be simplified as follows, when the register qubit is in $|0\rangle$, no operation is implemented as $e^{i0*At_0/2}=I$; when the register qubit is in $|1\rangle$, $e^{iAt_0/2}$ is applied to the first subsystem. After the phase estimation step, the state of the whole system is $\beta_1|u_1\rangle|0\rangle|0\rangle+\beta_2|u_2\rangle|1\rangle|0\rangle$. It can be seen that $0$ and $1$ of the register qubit corresponds to $\lambda_1$ and $\lambda_2$, respectively. But they each are only one bit of $\lambda_1$ and $\lambda_2$. Therefore, in general situations, a one-qubit register for quantum phase estimation is not enough to estimate the eigen values of  a $2\times 2$ matrix. Some prior knowledge of $\lambda_1$ and $\lambda_2$ is required for using only one qubit in the phase estimation process. We use the eigen values of 2 and 3 as an example. 2 has a binary form of 10. 3 has a binary form of 11. The two eigen values have a one-bit difference in their binary forms. Hence a one-qubit register for phase estimation can identify them, as long as we know their most significant bits are 1 in advance. We also mentioned that the controlled $R(\lambda^{-1})$ rotation is difficult to realize. However, when the register only has one qubit, it becomes easy. When the register qubit is $|0\rangle$ or $|1\rangle$, apply $2\sin^{-1}\frac{C}{\lambda_1}$ or $2\sin^{-1}\frac{C}{\lambda_2}$ to the third qubit respectively. Still using the eigen values  2 and 3 as an example, when the register qubit is in $|0\rangle$, we know this state corresponds to the eigen value 10, and the controlled $R(\lambda^{-1})$ rotation angle should be $2\sin^{-1}\frac{C}{2}$; when the register qubit is in $|1\rangle$, we know this state corresponds to the eigen value 11, and the controlled $R(\lambda^{-1})$ rotation angle should be $2\sin^{-1}\frac{C}{3}$. As concerns to the choice of $C$, on one hand we want the success probability $\Sigma_i|\frac{C\beta_i}{\lambda_i}|^2$ to be as large as possible, on the other hand, $C/\lambda_i$ should be reasonable sine values. For example, in the case with  $\lambda_1=2$ and $\lambda_2=3$, $C=2$ is a good choice. In this very simple three-qubit HHL, if the initial state of the third qubit is $|1\rangle$, the controlled $R(\lambda^{-1})$ rotation step can be replaced by a $|1\rangle$-controlled $R_y(\theta)$. When the register qubit is in $|0\rangle$, no operation is needed and when the register qubit is is $|1\rangle$, the  $R_y(\theta)$ rotation is applied to the third qubit, where $\theta=-2\cos^{-1}\frac{\lambda_1}{\lambda_2}$. From the point of view of experimental implementation, the $|1\rangle$-controlled $R_y(\theta)$ is a further simplification of the two state controlled $R(\lambda^{-1})$ rotation. The simplified three-qubit circuit is shown in Fig. \ref{HHL1}.

\subsection{Experimental implementation on Triangulum}

\begin{figure}[!ht]
\includegraphics[width=9cm]{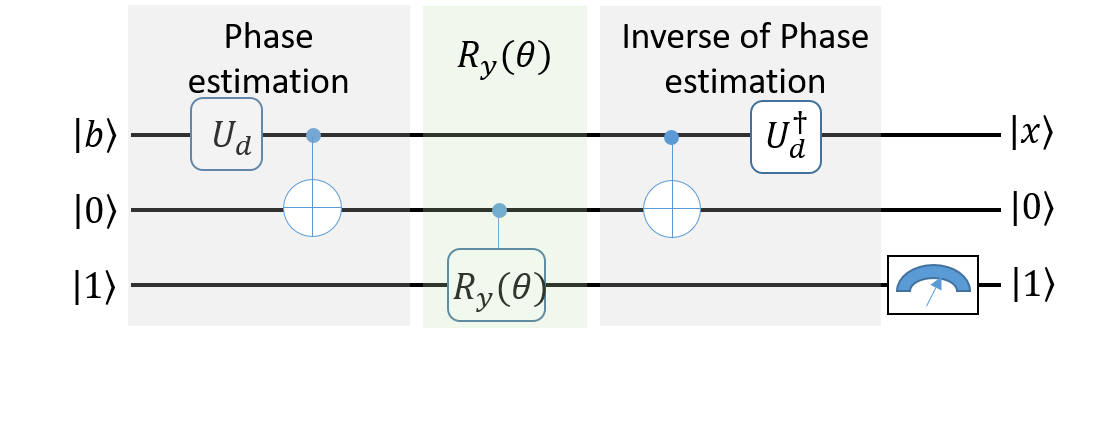}
\caption{The further simplified HHL circuit upon the knowledge of $U_d$.}
\label{HHL2}
\end{figure}

\begin{figure*}[!ht]
\includegraphics[width=17cm]{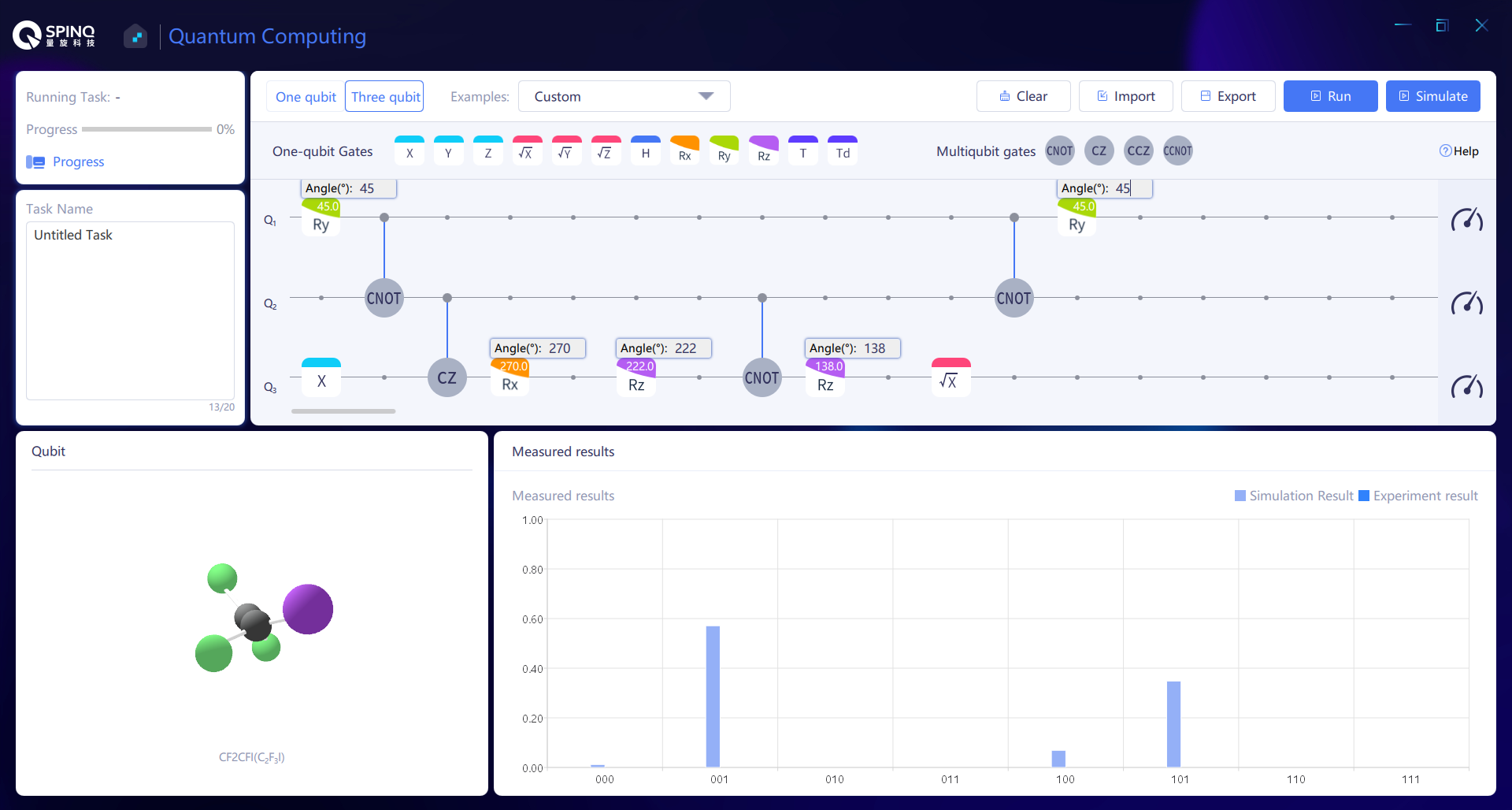}
\caption{The HHL sequence composed in \textsc{SpinQuasar}. }
\label{softHHL}
\end{figure*}

Here we realize the algorithm in the case
\begin{align}
&A=\begin{pmatrix}
2.14645 & -0.35355\\
-0.35355 & 2.85355\\
\end{pmatrix},
&b =\begin{pmatrix}
0.70711\\
0.70711\\
\end{pmatrix}.
\end{align} 
There are two two-qubit gates in the circuit. The $|1\rangle$-controlled $e^{iAt_0/2}$ gate can be decomposed to $U_d-{\rm CZ}-Ud^{\dagger}$ upon the knowledge of the diagonalization matrix of $A$, 
\begin{align}
A=U_d^{\dagger}\begin{pmatrix}
\lambda_1 & 0\\
0 & \lambda_2\\
\end{pmatrix}U_d.
\end{align}
Here $\rm CZ$ is the $|1\rangle$-controlled-$\sigma_z$ gate. $\rm CZ$ can be combined with the two $H$ gates and becomes the CNOT gate.  The $U_d^{\dagger}$ gate in the phase estimation step and the $U_d$ gate in the reverse of the phase estimation step can cancel each other, as shown in Fig. \ref{HHL2}. In the current case, $U_d=R_{-y}^1(\pi/4)$. Here, the superscription means the rotation is on the first qubit. The $|b\rangle$ state can be prepared using $R_y^1(\frac{\pi}{2})$. The $|1\rangle$-controlled $R_y(\theta)$ operation can be realized in different ways. Here, we decompose it as
\begin{align}
{\rm CZ}&-R_{-x}^3(\frac{\pi}{2})-R_{-z}^{3}(\frac{\pi-\theta}{2})\nonumber\\&-{\rm CNOT}-R_z^3(\frac{\pi-\theta}{2})-R_x^3(\frac{\pi}{2}).
\end{align} 
Now, the required operations of the HHL algorithm are decomposed to basic gates that can be realized by Triangulum, and the complete gate sequence is as follows
\begin{align}
&R_x^3(\pi)-R_y^1(\frac{\pi}{4})-{\rm CNOT_{12}}-{\rm CZ_{23}}-R_{-x}^3(\frac{\pi}{2})-R_{-z}^3(\frac{\pi-\theta}{2})\nonumber\\-&{\rm CNOT_{23}}-R_z^3(\frac{\pi-\theta}{2})-R_x^3(\frac{\pi}{2})-{\rm CNOT_{12}}-R_y^1(\frac{\pi}{4})
\end{align}
The first gate in the above sequence prepares the third qubit in the state $|1\rangle$. It should be mentioned that the gate used to prepare $|b\rangle$ is combined with the gate $U_d$ and hence only one gate $R_y^1(\frac{\pi}{4})$ is implemented in front of the first $\rm CNOT_{12}$ gate. The construction of this sequence using \textsc{SpinQuasar} is illustrated in Fig. \ref{softHHL}.

\begin{figure}[!ht]
\includegraphics[width=9cm]{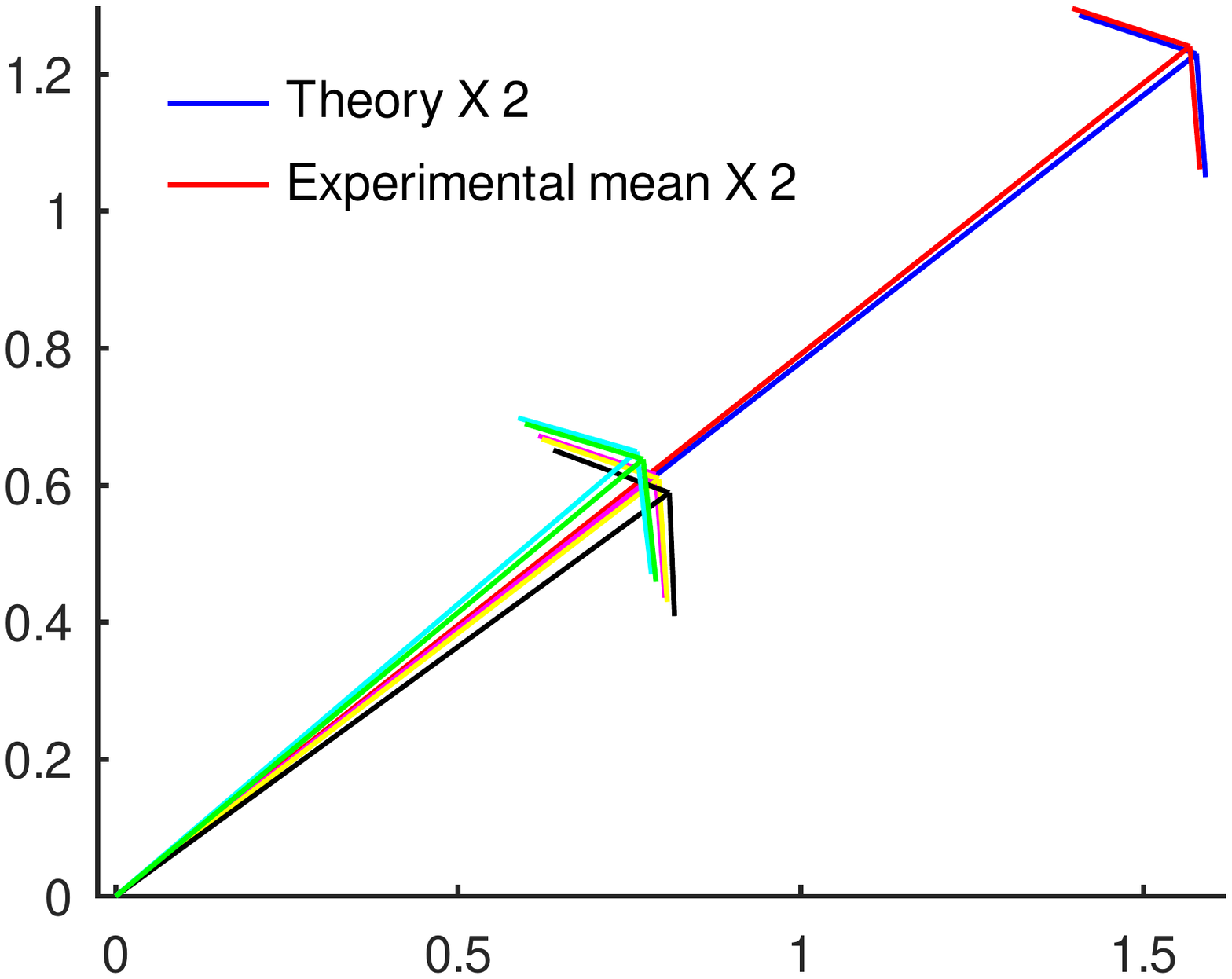}
\caption{The experimental results of the five repetitions are shown as the vectors. The mean vector of the five repetitions are also shown in red to compare with the theoretical vector in blue. Both the mean vector and the theoretical vector are multiplied by 2 for clarity.}
\label{results}
\end{figure}

Triangulum measures the probabilities of all the eight basis states $\rho_{ii},i=1\ldots 8$, as illustrated in Fig. \ref{softHHL}. Here the subscription $i=1\ldots 8$ corresponds to the base $|000\rangle, |001\rangle, |010\rangle, |011\rangle, |100\rangle, |101\rangle, |110\rangle, |111\rangle$. From the analysis above, we know  when the third qubit is found in $|1\rangle$, the first qubit is in the state $|x\rangle$.  Since ideally the final state of the second qubit is $|0\rangle$, we use $\rho_{22}$ and $\rho_{66}$, which are the probabilities of $|001\rangle$ and $|101\rangle$, to infer $|x\rangle$, 
\begin{align} 
&|x\rangle\propto\begin{pmatrix}
\sqrt{\rho_{22}}\\
{\bf s}\sqrt{\rho_{66}}\\
\end{pmatrix}
\end{align}
The result can be directly written as the square root of the probabilities is because $A$ and $\vec{b}$ are both real and hence $\vec{x}$ is real as well. ${\bf s}$ is the relative sign between the two entries of $\vec{x}$, or $|x\rangle$, which is also the sign of the coherence term between   $|001\rangle$ and $|101\rangle$, $\rho_{26}$, in the density matrix. This coherence term can be mapped to the measurable probabilities by a gate $R_{-y}^1(\frac{\pi}{2})$. After this gate, if $\rho_{22}>\rho_{66}$, then $\bf s=1$, otherwise $\bf s=-1$. Five repetition of the experiments are done. The result vectors are shown in Fig. \ref{results}. The mean vector of the results is also shown and compared with the theoretical vector, which are only $\sim 0.4^\circ$ apart. And their tangent values have a difference of only about 1.5 percent. The good agreement between experimental and theoretical results implies a successful proof of principle demonstration of the HHL algorithm.

\section{Conclusion}
The successful demonstration of the HHL algorithm shows Triangulum's great potential both in quantum information education and quantum computing research. The embedded two-qubit and three-qubit quantum algorithms provides great examples for quantum computing learners. The powerful advanced pulse control functions, which provides arbitrary waveform generation ability, is a great asset for research in quantum control in realistic environments for advanced users. SpinQ will continue to develop desktop quantum computing platform with more qubits. Meanwhile, a simplified version of SpinQ Gemini, namely Gemini Mini \cite{geminiMinilink}, has been recently realised. Gemini Mini is much more portable ($20\times 35
\times 26$ cm$^3$, $14$ kg) and affordable for most K-12 schools around the world.

\bibliography{trRef}

\end{document}